\documentclass[aps,pra,reprint,superscriptaddress]{revtex4-2}
\usepackage{graphicx}
\usepackage{dcolumn}
\usepackage{bm}
\usepackage[T1]{fontenc}
\usepackage{newtxtext,newtxmath}
\usepackage{amsmath}
\usepackage{nicefrac}
\usepackage{ulem}
\usepackage{color}
\usepackage{xcolor}
\usepackage{microtype}
\usepackage[colorlinks=true,linkcolor=blue,citecolor=blue]{hyperref}
\usepackage{enumerate}
\usepackage{enumitem}
\usepackage{gensymb}

\newcommand{\ikf}{Institut f\"ur Kernphysik, Goethe-Universit\"at Frankfurt, Max-von-Laue-Str. 1, 60438 Frankfurt am Main, Germany}
\newcommand{\ESRF}{ESRF, The European Synchrotron, 71 Avenue des Martyrs, CS 40220, 38043 Grenoble Cedex 9, France}
\newcommand{\kassel}{Institut f\"ur Physik und CINSaT, Universit\"at Kassel, Heinrich-Plett-Str. 40, 34132 Kassel, Germany}
\newcommand{\berlin}{Molecular Physics, Fritz-Haber-Institut der Max-Planck-Gesellschaft, Faradayweg 4-6, 14195 Berlin, Germany}
\newcommand{\HD}{Max-Planck-Institut f\"ur Kernphysik, Saupfercheckweg 1, 69117 Heidelberg, Germany}
\newcommand{\EXFEL}{European XFEL, Holzkoppel 4, 22869 Schenefeld, Germany}
\newcommand{\zhejiang}{School of Physics, Zhejiang Key Laboratory of Micro-Nano Quantum Chips and Quantum Control, Zhejiang University, Hangzhou 310058, China}
\newcommand{\hannover}{Institut f\"ur Theoretische Physik, Leibniz Universit\"at Hannover, Appelstr. 2, 30167 Hannover, Germany}

\newcommand{\eCompton}{\ensuremath{e^-}}
\newcommand{\eAuger}{\ensuremath{e^-_{\text{A}}}}
\newcommand{\Ne}{\ensuremath{\text{Ne}}}
\newcommand{\Oxy}{\ensuremath{\text{O}}}
\newcommand{\Car}{\ensuremath{\text{C}}}
\newcommand{\photonIn}{\ensuremath{\gamma}}
\newcommand{\photonOut}{\ensuremath{\gamma'}}

\newcommand{\eComptonenergy}{\ensuremath{E_e}}
\newcommand{\eComptonEfree}{\ensuremath{E_{e}^{\mathrm{free}}}}
\newcommand{\eComptonVec}{\ensuremath{\vec{k}_{e}}}
\newcommand{\eComptonMag}{\ensuremath{k_{e}}}

\newcommand{\eAugerVec}{\ensuremath{\vec{k}_{e\mathrm{A}}}}

\newcommand{\RecVec}{\ensuremath{\vec{K}_{\mathrm{i}}}}
\newcommand{\NeVec}{\ensuremath{\vec{K}_{\mathrm{Ne}}}}

\newcommand{\CVec}{\ensuremath{\vec{K}_{\mathrm{C}}}}

\newcommand{\OVec}{\ensuremath{\vec{K}_{\mathrm{O}}}}

\newcommand{\photonscatang}{\ensuremath{\theta_{\gamma}}}
\newcommand{\photonInE}{\ensuremath{E_{\gamma}}}
\newcommand{\photonInVec}{\ensuremath{\vec{k}_{\gamma}}}
\newcommand{\photonInMag}{\ensuremath{{k}_{\gamma}}}
\newcommand{\photonOutE}{\ensuremath{E_{\gamma'}}}
\newcommand{\photonOutVec}{\ensuremath{\vec{k}_{\gamma'}}}
\newcommand{\photonOutMag}{\ensuremath{{k}_{\gamma'}}}
\newcommand{\ETrans}{\ensuremath{\Delta E}}
\newcommand{\MomTransVec}{\ensuremath{\vec{Q}}}
\newcommand{\MomTransMag}{\ensuremath{Q}}
\newcommand{\MomTransAngle}{\ensuremath{\theta_{Q}}}
\newcommand{\Ebind}{\ensuremath{I_P}}
\newcommand{\eComptonEbound}{\ensuremath{E_{e}^{\mathrm{bound}}}}

\begin{document}
DOI: 10.1103/bcl4-3cnr\newline
Phys.~Rev.~A~114,~023105,~2026\newline
Author's version
\title {
\large
The influence of the binding energy on Compton-scattering-driven ionization}
\author{N.~Melzer} \email{melzer@atom.uni-frankfurt.de} \affiliation{\ikf}
\author{M.~Kircher} \affiliation{\ikf}
\author{G.~Kastirke} \affiliation{\ikf}
\author{J.~Rist} \affiliation{\ikf}
\author{D.~Tsitsonis} \affiliation{\ikf}
\author{A.~Pier} \affiliation{\ikf}
\author{L.~Kaiser} \affiliation{\ikf}
\author{J.~Kruse} \affiliation{\ikf}
\author{N.~Anders} \affiliation{\ikf}
\author{J.~Stindl} \affiliation{\ikf}
\author{L.~Sommerlad} \affiliation{\ikf}
\author{O.~D.~McGinnis} \affiliation{\ikf}
\author{M.~Schmidt} \affiliation{\ikf}
\author{L.~Nowak} \affiliation{\ikf}
\author{A.~Kügler} \affiliation{\ikf}
\author{I.~Dwojak} \affiliation{\ikf}
\author{P.~Daum} \affiliation{\ikf}
\author{P.~Roth} \affiliation{\ikf}
\author{M.~Astaschov} \affiliation{\ikf}
\author{J.~Drnec} \affiliation{\ESRF}
\author{K.~Lin} \affiliation{\ikf}\affiliation{\zhejiang}
\author{N.~Wong} \affiliation{\ikf}
\author{F.~Trinter} \affiliation{\berlin}
\author{M.~S.~Sch\"offler} \affiliation{\ikf}
\author{L.~Ph.~H.~Schmidt} \affiliation{\ikf}
\author{D.~M.~Haubenreißer} \affiliation{\kassel}
\author{N.~M.~Novikovskiy} \affiliation{\kassel}
\author{N.~Diehl} \affiliation{\kassel}
\author{Ph.~V.~Demekhin} \affiliation{\kassel}
\author{S.~Brennecke} \affiliation{\hannover}
\author{M.~Lein} \affiliation{\hannover}
\author{S.~Eckart} \affiliation{\ikf}
\author{T.~Jahnke} \affiliation{\HD}\affiliation{\EXFEL}
\author{R.~D\"orner} \email{doerner@atom.uni-frankfurt.de} \affiliation{\ikf}

\begin{abstract}
\noindent
We present a comprehensive study of Compton scattering of 20~keV photons by bound electrons over a wide range of binding energies (Ne~2p, Ne~2s, C~1s, O~1s, and Ne~1s orbitals). In particular, we have measured fully differential cross sections, i.e., electron angular distributions with respect to the direction of the photon momentum transfer for fixed electron energy. Our data show strong deviations from the predictions of the ``quasi-free'' electron approximation. The binding energy and the shape of the Coulomb potential from which the electron escapes both play a significant role in its kinematics. We find that Compton electrons are scattered and even back-reflected by the ionic potential. The strength of this effect depends on the orbital from which the electron is ejected.
\end{abstract}
\maketitle

\section{Introduction}
\label{secintro}

When light interacts with matter, Compton scattering can occur as a fundamental process. Upon its discovery \cite{AHCompton1922}, it was described as a binary collision of a photon with a ``quasi-free'' electron. In this approximation, the momentum vector of the Compton electron is linked to the scattering angle of the photon (\photonscatang), and the energy loss of the photon equals the electron's kinetic energy, which is given by (atomic units are used throughout)

\begin{align}
\eComptonEfree\left(\photonscatang\right) = \photonInE - \photonOutE = \photonInE - \frac{\photonInE}{1 + \nicefrac{\photonInE}{c^2} \left(1 - \cos\photonscatang\right)} \ .
\label{eqnEfree}
\end{align}

Under these free conditions, the cross section is to a good approximation given by the Klein-Nishina equation \cite{KleinNishina1929}. In more realistic scenarios, however, the electron from which the photon scatters is bound. For Compton scattering, this has rich consequences, such as:

\begin{enumerate}[label=(\alph*), leftmargin=2em]
\item The electron is not at rest but has a momentum distribution given by the square of the Fourier transform of its bound-state wave function.
\item In all real neutral systems except hydrogenic atoms, there is more than one electron, and the single-active-electron approximation might not hold.
\item Some energy is needed to overcome the binding energy, which changes the kinematics of the process.
\item The electron will be influenced by the potential from which it escapes. This potential can even be non-spherically symmetric, as is the case for Compton scattering in molecules \cite{Haubenreisser2025,Haubenreisser2025chiral}.
\end{enumerate}

With respect to (a), the role of the initial-state momentum distribution has already been recognized in very early studies \cite{DuMond1929}. For cases of high photon energies, where the magnitude of the momentum transfer from the photon to the electron is much larger than the magnitude of the electron momenta in the bound state, the initial electron momentum vector is added to the momentum balance. This is well understood today and is accounted for by the impulse approximation \cite{Cooperbook}. It makes Compton scattering a powerful tool for measuring momentum distributions in bound orbitals, which are therefore called \textit{Compton profiles}.

Considering (b), many-electron effects have been studied at the level of total cross sections for helium in the 1990s \cite{Spielberger1999} and more recently in 2021 \cite{Chuluunbaatar2021272}. A prominent consequence of these effects is the double ionization of helium induced by Compton scattering \cite{Chuluunbaatar2021278}. As in the case of photoionization, the double-ionization probability by Compton scattering is in the few-percent range, and two mechanisms -- shake-off and knock-off -- have been identified to facilitate it. The former is a consequence of correlation in the initial state, and the latter is mediated by electron-electron interaction after the Compton scattering. A rigorous treatment involving correlation in the initial state can be found in Refs.~\cite{Kaplan2003,BERGSTROM19973}.

Finally, consequences (c) and (d), i.e., the roles of the binding energy and the Coulomb potential, are the topic of the present paper. One known effect of considering a bound rather than a free electron is a reduction of the total Compton cross section at low photon energies when
$\eComptonEfree\left(\photonscatang\right)$ becomes smaller than the binding energy for an increasing range of photon scattering angles \cite{Samson1994}. However, on a more detailed level, the influence of the binding energy on Compton scattering has been rarely studied. For example, how does the binding change the electron emission pattern, and how are fully differential cross sections influenced by the binding? To tackle these questions experimentally in a systematic fashion, we compare Compton scattering of 20~keV photons by electrons from the Ne L shell (48.5~eV binding energy for Ne 2s or 21.6~eV for Ne 2p) with Compton scattering at the Ne K shell (870.2~eV binding energy) \cite{xraydatabooklet,Cardona2005}. Further, we include data on Compton scattering at the C (296.2~eV binding energy) and O (542.5~eV binding energy) K shells in CO \cite{MYRSETH200257, jahnke2011Ckshell, WANG2021147103}. The expectation is that for the Ne L shell, deviations from the predictions of the impulse approximation will be smaller, while they will become massive for K-shell ionization. The goal of the paper is to elucidate Compton scattering from all possible perspectives and at all levels of detail.

\begin{figure}[!t]
\center
\includegraphics[width=1.\columnwidth]{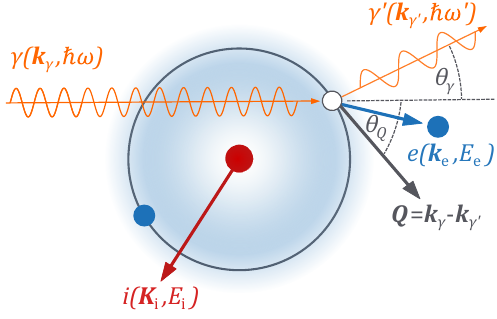}
\caption{Sketch of the kinematics of non-relativistic Compton scattering. A photon with incoming momentum vector \photonInVec{} is scattered to a final state with momentum \photonOutVec{}, transferring a momentum $\MomTransVec = \photonInVec - \photonOutVec$ and an energy $\ETrans = c \photonInMag - c \photonOutMag$. The momentum transfer is shared between the Compton electron (\eComptonVec) and the ion (\RecVec).}
\label{Fig1}
\end{figure}

Single ionization by Compton scattering involves three particles: the photon, which has transferred a momentum \MomTransVec{} to the atom or molecule at which it is scattered, the momentum \eComptonVec{} of the ejected Compton electron, and the momentum \RecVec{} of the recoiling ion (see Fig.~\ref{Fig1} for a sketch). The nine momentum components are constrained by momentum and energy conservation, leaving five independent observables and giving rise to a five-fold differential cross section. We will not consider the dependence of Compton scattering on the polarization of the photon, since this factorizes \cite{Jauch1955, Boca_2015}. Neglecting the polarization leaves four independent observables, making the cross section for Compton scattering four-fold differential. ``Four-fold'' means that the cross section is fully described by any set of four independent observables. We choose, e.g., the photon scattering angle, the Compton electron polar and azimuthal angles with respect to the light propagation direction, and the electron energy ($\photonscatang$, $\theta_e$, $\phi_e$, $E_e$), which yields the cross section

\begin{eqnarray}
\label{eqnfourcrosssecspher}
\frac{\mathrm{d}^4\sigma}{\mathrm{d}\theta_e \mathrm{d}\phi_e \mathrm{d}E_e \mathrm{d}\photonscatang} .
\end{eqnarray}

Another choice of observables, which is well adapted to textbook figures of Compton scattering (Fig.~\ref{Fig1}), is the angle of the momentum transfer \MomTransAngle{} and the Cartesian momentum coordinates of the electron ($\MomTransAngle$, $k_{e,x}$, $k_{e,y}$, $k_{e,z}$), i.e., the cross section

\begin{eqnarray}
\label{eqnfourcrossseckart}
\frac{\mathrm{d}^4\sigma}{\mathrm{d}k_{e,x}\mathrm{d}k_{e,y}\mathrm{d}k_{e,z}\mathrm{d}\MomTransAngle} .
\end{eqnarray}

Here, we exploit the power of the Cold Target Recoil Ion Momentum Spectroscopy (COLTRIMS) reaction microscope, i.e., its ability to cover the full final-state phase space of all particles in a single measurement. This feature allows us to freely choose coordinate frames and observables adapted to the physical question of interest. Accordingly, we divide this paper into three sections, starting with two-fold differential cross sections in Section~\ref{sectwofold}, and building up to Section~\ref{secfourfold}, which contains the most detailed information in fully differential cross sections.

We compare our data to the predictions of the $A^2$ approximation \cite{Pisk_2016,BERGSTROM19973}. In this approximation, ionization by Compton scattering is described by two independent steps that are linked only via energy and momentum conservation. One step is the scattering of the photon by an electron, as given by the Klein-Nishina cross section \cite{KleinNishina1929,Pisk_2016}

\begin{eqnarray}
\label{eqnkn}
\left( \frac{\mathrm{d}\sigma}{\mathrm{d}\Omega} \right)^{\text{KN}} = \frac{r_0^2}{2} \left(\frac{\omega'}{\omega}\right)^2 \left( \frac{\omega}{\omega'} + \frac{\omega'}{\omega} - \sin^2 \photonscatang \right) ,
\end{eqnarray}

where $r_0$ is the classical electron radius. The other step is the emission of the electron as a consequence of the momentum transfer in the photon-scattering step. This second step is described by the overlap integral of the ground state -- shifted in momentum space by the momentum transfer -- with the continuum state for the given energy. For high photon energies and large scattering angles, this converges to the impulse approximation, in which the escaping electron is approximated by a plane wave. Thus, in the $A^2$ approximation, the single differential cross section $\mathrm{d}\sigma\left(\photonscatang\right)/\mathrm{d}\photonscatang$ is the product of the Klein-Nishina cross section and the incoherent scattering factor \cite{BERGSTROM19973} for the momentum \MomTransMag{} in the quasi-free case $\MomTransMag{}_{\text{free}}\left(\photonscatang\right)= \sqrt{2\Delta E}$, with $\Delta E$ given by Eq.~(\ref{eqnEfree}).

\begin{figure*}
\center
\includegraphics[width=0.8\linewidth]{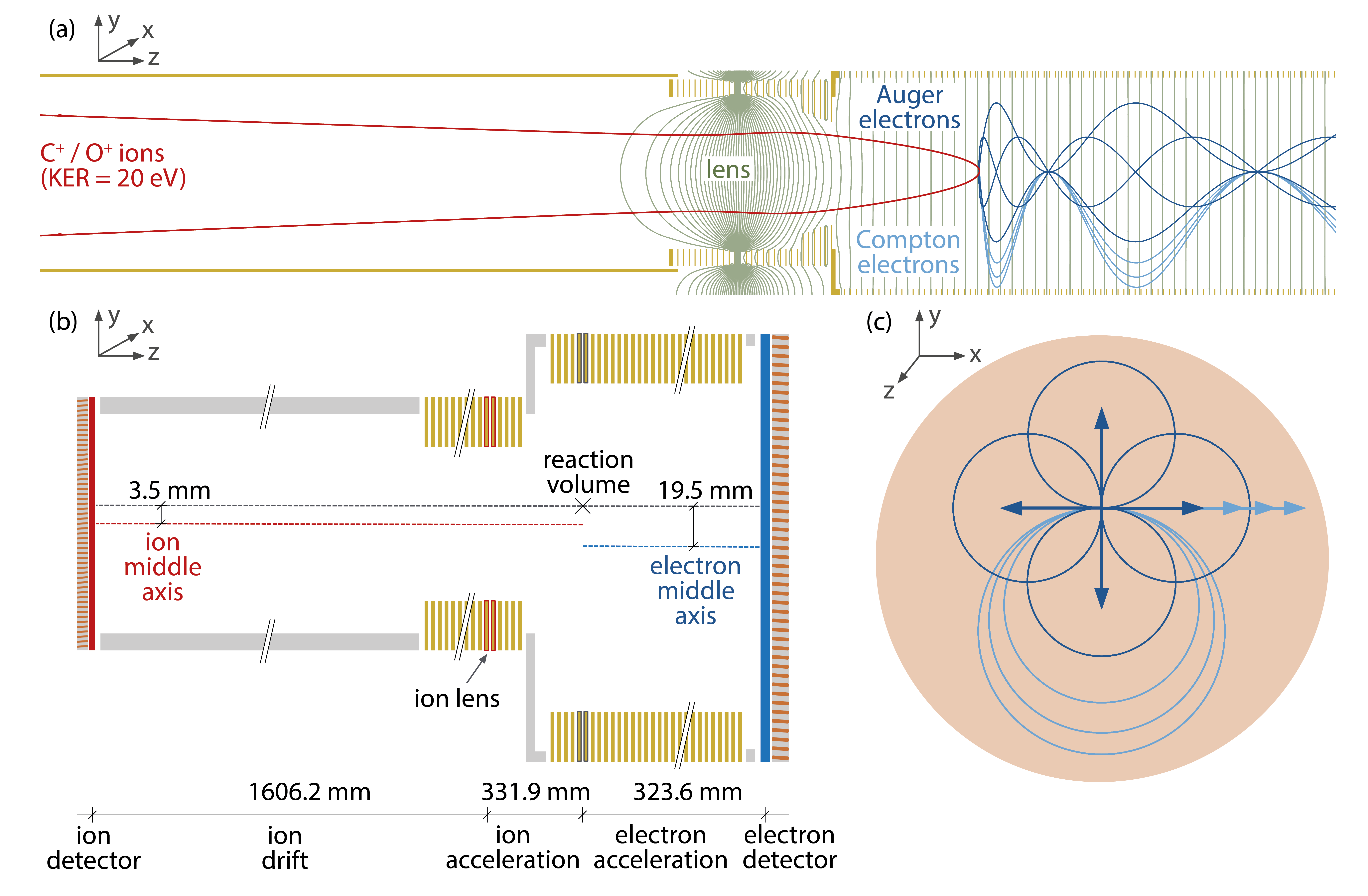}
\caption{Simulation and sketch of the COLTRIMS spectrometer used in the experiment (not to scale). The $x$ axis is defined as the photon propagation direction, the $y$ axis as the target propagation direction, and the $z$ axis as the time-of-flight direction. (a) Simulation of the ion (red) and electron (blue) trajectories inside the electromagnetic field (green) in the $yz$ plane. C$^+$ and O$^+$ ions dissociate with a KER of 20~eV. An electromagnetic lens is used to focus the ions onto the detector. Compton electrons are shown with energies of 1500, 2000, and 2400~eV. Auger electrons have an energy of 780~eV. For the simulation, example trajectories with maximal energies are shown to demonstrate the size of the spectrometer (C$^+$ and O$^+$ ions on the ion side and Auger electrons from Ne on the electron side). (b) Electrons pass through an acceleration area onto a 150~mm diameter electron detector. The reaction volume is shifted 19.5~mm compared to the middle axis of the electron detector in order to measure Auger and forward-directed Compton electrons up to 2~keV. Besides the acceleration area, a combination of lens and ion drift zone is implemented on the ion side of the spectrometer to achieve an ion momentum resolution of about 0.13~a.u. The sketch is not to scale, e.g. in panel (b) the $z$ direction is compressed. (c) Simulated electron trajectories shown in the $xy$ plane. High-energy Compton electrons are emitted in the photon propagation direction (positive $x$ direction). Due to the magnetic field, they hit the detector displaced in the negative $y$ direction, which is why the detector is shifted in $y$.}
\label{Fig2}
\end{figure*}
The present paper builds on previous studies by our group \cite{Kircher2020,Kircher2022,Melzer2024}, in which three aspects of Compton scattering were introduced. We elaborate on these in more detail and present additional data in the current paper. The first aspect is the role of the binding energy. Previously, we studied a limiting case by measuring Compton scattering on He employing a photon energy of 2~keV \cite{Kircher2020}. At this low energy, the probability of the Compton effect is much reduced, as emphasized by the fact that the energy transfer corresponding to backscattering of a 2~keV photon from a free electron at rest is smaller than the binding energy of He. The second aspect is the extraction of the initial-state Compton profile by measuring the ion momenta. We studied this for He single and double ionization in Ref.~\cite{Kircher2022}. In the present paper, we extend this study to Ne. The third aspect is the scattering of the Compton electron at the target nucleus, which we investigated in Ref.~\cite{Melzer2024} for the Ne L shell. We expand on these studies in the present paper by also including Compton scattering at K-shell electrons of Ne and at the C and O atoms of CO molecules.

\section{Experiment}
\label{secexperiment}

Coincidence experiments on Compton scattering are highly challenging. In order to extract low-energy electrons from the scattering target in an interaction-free manner, such experiments have to be performed in the gas phase. The small Compton cross section (10$^{-24}$~cm$^2$) and the dilute gas sample (about $1$~mm wide jet, $10^{11}$~particles/cm$^2$ target density for, e.g., Ne) result in an expected rate of only 25~Hz Compton events, even when considering the most brilliant synchrotron beams available with a flux of $2.6 \times 10^{14}$~photons/s. These numbers demonstrate that a fully differential experiment is impossible with detectors that cover only a small fraction of the full $4\pi$ solid angle. This precludes today the use of photon-electron coincidence schemes (as were used in the only coincidence experiment on solid-state targets \cite{Cooper1985,Bell1991,Metz1999}) in experiments employing gas targets.

There is, however, an alternative that avoids the necessity of detecting the scattered photon directly while still providing access to fully differential cross sections. If the momenta of the Compton electron and the ion are measured in coincidence, the energy transfer \ETrans{} and momentum transfer \MomTransVec{} between the photon and the target can be calculated using momentum and energy conservation. This enables a full $4\pi$ solid-angle coverage of the Compton process in the laboratory frame. Additionally, this approach has the decisive experimental advantage that the energy and momentum resolution are not hampered by the bandwidth of the photon. In photon-detection schemes, the energy and momentum transfer are obtained as the difference of two large quantities, \photonInVec{} and \photonOutVec. Thus, the experiment requires a narrow-bandwidth incoming photon beam. On the contrary, our alternative approach determines \ETrans{} and \MomTransVec{} from the electron and ion momenta with a precision that is almost independent of the incoming photon energy resolution. This allows us to use so-called ``pink'' beams (i.e., unmonochromatized undulator radiation) or broad-bandwidth photons from the synchrotron, resulting in a much higher photon flux compensating for the small Compton cross section.

\begin{figure}[!t]
\center
\includegraphics{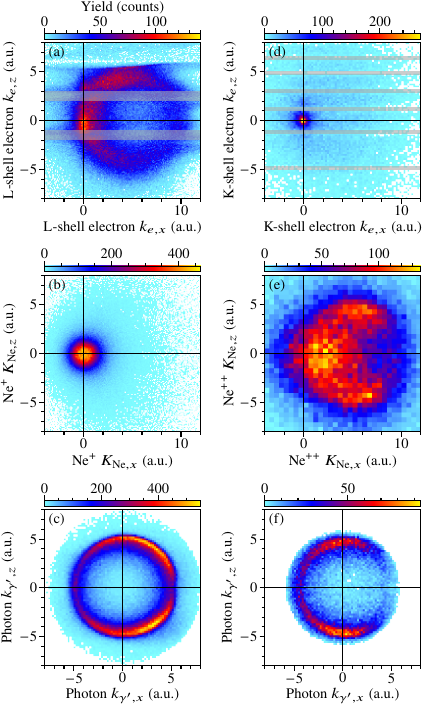}
\caption{Electron, ion, and photon momentum distributions for Compton scattering of 20~keV photons at the Ne L (a--c) and K (d--f) shells. (a,d) Measured L-shell [K-shell] electron momentum distribution in the plane defined by the incoming photon (horizontal) and the time-of-flight direction of the spectrometer (vertical). The gray bars indicate regions where the time-of-flight is an integer multiple of the gyration period in the magnetic guiding field (see Ref.~\cite{Ullrich2003} and main text for explanation). The third momentum component is integrated over. (b,e) Momentum distribution of Ne$^+$ [Ne$^{++}$] recoiling ions measured in coincidence with the electrons shown in (a) [(d)]. For the Ne$^{++}$ ions [panel (e)], Auger electron momenta are added. (c,f) Momentum distribution of the scattered photon calculated event-by-event from the measured ion and electron momenta, assuming an incoming photon energy
of 20~keV.}
\label{Fig3}
\end{figure}

In more detail, we measured electrons with energies up to 2~keV and ions with a kinetic energy release (KER) up to 20~eV with a collection angle of $4\pi$ using a COLTRIMS reaction microscope \cite{Dorner2000,Ullrich2003}. The experimental setup (Fig.~\ref{Fig2}) used to record the presented data is very similar to the one explained in Ref.~\cite{Melzer2024}. At beamline ID31 at the European Synchrotron Radiation Facility (ESRF), X-ray transfocators \cite{Vaughan2011} were utilized to provide photons with a bandwidth of about 2\%. The linearly polarized beam with an energy of $\photonInE = 20$~keV ($\photonInMag = 5.37$~a.u.) and an intensity of around $2.6 \times 10^{14}$~photons/s was crossed with a supersonic gas jet of Ne and CO target atoms/molecules. Parallel 31~V/cm electric and 35~Gauss magnetic fields were employed to guide electrons and ions with energies below 2~keV onto two time- and position-sensitive detectors with hexagonal delay-line readout \cite{Jagutzki2002}. To capture forward-emitted high-energy Compton electrons, the electron detector was placed 19.5~mm off-center and had an active diameter of 150~mm.
%
%
%
The ion arm had a total length of 1938~mm and incorporated an electrostatic lens \cite{Dorner1997}. The reaction volume has a size of about $1.0\times 0.4\times 0.1$~mm ($x, y, z$). The finite reaction volume corresponds to an electron and ion momentum resolution of about 0.4 and 0.13~a.u., respectively. This systematic error was estimated to be negligible compared to the statistical errors.
The single-electron count rate was 1.3~kHz, the single-ion count rate was 1.9~kHz, and the coincidence rate was 25~Hz. The local target pressure at interaction point was about 10$^{-5}$~mbar, and the vacuum in the main chamber where the target zone is located and the reaction takes place was as low as $2 \times 10^{-9}$~mbar.

As indicated above, neon and carbon monoxide served as targets in our experiments. The following three reaction channels were measured:

\begin{eqnarray}
    \photonIn(20~\text{keV}) + \Ne &\rightarrow& \Ne^+(2\text{s},2\text{p})^{-1} + \eCompton + \photonOut \label{eqL} \\ 
    \photonIn(20~\text{keV}) + \Ne &\rightarrow& \Ne^+(1\text{s})^{-1} + \eCompton + \photonOut \label{eqK1} \\ 
                 &          & \downarrow                              \nonumber \\
                 &         &\Ne^{2+} + \eAuger + \eCompton + \photonOut \label{eqK2} \\
    \photonIn(20~\text{keV}) + \Car\Oxy &\rightarrow& \Car^{+} + \Oxy^{+} + \eAuger + \eCompton + \photonOut \label{eqK3}
\end{eqnarray}

Correspondingly, the respective momentum balances for Eqs.~(\ref{eqL}), (\ref{eqK2}), and (\ref{eqK3}) read:

\begin{eqnarray}
\begin{split}
    \photonInVec &=& \NeVec + \eComptonVec + \photonOutVec \\
    \photonInVec &=& \NeVec + \eAugerVec + \eComptonVec + \photonOutVec \\
    \photonInVec &=& \CVec + \OVec + \eAugerVec + \eComptonVec + \photonOutVec
    \end{split}
\end{eqnarray}

The momenta \eAugerVec{} and \eComptonVec{} refer to Auger and Compton electrons, which were both detected in our experiment in coincidence. The subscripts C and O in \CVec{} and \OVec{} denote the measured C and O ions.

Figure \ref{Fig3} shows electron, ion, and photon momentum distributions for Compton scattering at the Ne L [panels (a-c)] and K [panels (d-f)] shells [Eqs.~(\ref{eqL}) and (\ref{eqK2})]. The Compton electrons in panels (a) and (d) are presented in the plane defined horizontally by the incoming photon ($x$ axis) and vertically by the spectrometer axis ($z$ axis). The photon polarization vector is parallel to the $y$ axis (see Fig.~\ref{Fig2}).

{\color{orange}
}

When detecting electrons using a COLTRIMS reaction microscope, certain restrictions arising from the employed magnetic guiding field need to be considered. In cases where the electron time-of-flight is an integer multiple of the electron's gyration period in the magnetic field, the momentum of the electron cannot be resolved (see Ref.~\cite{Ullrich2003}). We have therefore set up the magnetic field in a way that the resulting blind spots [corresponding to the gray areas in Figs.~\ref{Fig3}(a) and \ref{Fig3}(d)] are distributed asymmetrically with respect to $\eComptonMag{}_{,z}=0$. Due to symmetry reasons, the information within the gray areas is still available in our measurement, as events that yielded a corresponding $\eComptonMag{}_{,z}$ with opposite sign can be inverted and used to partly fill the grey areas. In most cases, however, the results in all other figures except Fig.~\ref{Fig3} are not significantly affected by this, as they do not depict the laboratory frame directly.

In addition to the Compton ionization process, photoabsorption also occurs (with a cross section 60 times larger than Compton scattering). Events from photoabsorption can be clearly separated from those from Compton scattering by the large electron and ion momenta (see, e.g., Ref.~\cite{Spielberger95} and Fig.~1 in Ref.~\cite{Schmidt2024}). The photoelectron recoil momentum is about 37.5~a.u., compared to the ions resulting from Compton ionization with momenta below 12~a.u. [Figs.~\ref{Fig3}(b) and \ref{Fig3}(e)]. Events from Compton scattering at the L shell [Eq.~(\ref{eqL})] can be distinguished from events from Compton scattering at the K shell [Eqs.~(\ref{eqK1}) and (\ref{eqK2})] through the coincident detection of doubly charged ions together with the detection of a Compton and one Auger electron. Compton and Auger electrons are distinguished by energy: Auger electrons have characteristic energies of 780~eV (Ne~1s), 255~eV (C~1s), and 500~eV (O~1s). This separation is almost unambiguous for Auger electrons emitted in the backward direction because high-energy Compton electrons are mostly forward-emitted.

In Fig.~\ref{Fig3}(c) and (f) the momentum distributions of the scattered photon are shown. The incoming photon energy is $20$~keV and the distributions are calculated -- as explained above -- event-by-event from the measured ion and electron momenta. There are two major differences between the scattered photon distribution for Ne L- [panel (c)] and K-shell [panel (f)] ionization: First, the radius of the ring in panel (f) is smaller than in panel (c), reflecting the higher binding energy, which reduces the maximum energy of the scattered photon. Second, the distribution of events on the ring is more enhanced in the forward direction (positive $k_{\gamma,x}$) in panel (c), while forward-scattered photons are missing in panel (f). For L-shell ionization, the distribution is close to the one expected by the Klein-Nishina cross section. For K-shell ionization, however, $\eComptonEfree$ is smaller than the binding energy for cases of a scattering into the forward hemisphere, which suppresses the Compton process.

\section{Two-fold differential cross sections}
\label{sectwofold}

\begin{figure}[!t]
\center
\includegraphics{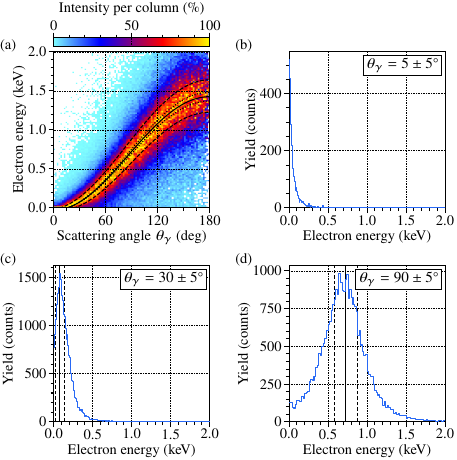}
\caption{Double differential cross section $\mathrm{d}^2\sigma/\mathrm{d}\photonscatang \mathrm{d}\eComptonenergy{}$ for Ne L-shell ionization by Compton scattering of a 20~keV photon. (a) Horizontal axis: photon scattering angle \photonscatang, vertical axis: electron energy, with column normalized counts to highlight the overall distribution. (b)-(d) Projections from (a) onto the vertical axis for the angular regions indicated in the graphs. The black solid curves show Eq.~(\ref{eq:Eebound}), i.e., the locus of events for a binary encounter with an electron at rest, where the average binding energy of \Ebind{} $= (2 \times 48.5 + 6 \times 21.6) / 8$~eV is subtracted from the electron energy after the kick.
The dotted curves in panel (a) correspond to a $2$\% bandwidth of the incoming photon energy.
The dashed curves show the corresponding energies for the case in which the electron was initially not at rest, but had a momentum distribution given by the Ne L shell (see main text for details).}
\label{Fig4}
\end{figure}

\begin{figure}[!b]
\center
\includegraphics{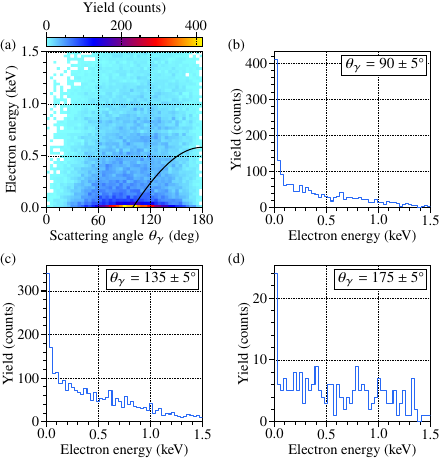}
\caption{Analogous to Fig.~\ref{Fig4}, but for K-shell ionization ($I_P = 870.2$~eV). Panel (a) shows the absolute yield, and panels (b)-(d) show the projections for different scattering angles (see insets).}
\label{Fig5}
\end{figure}
We first examine the doubly differential cross section in terms of the photon scattering angle $\theta_\gamma$ and the resulting Compton electron energy:

\begin{align}
\frac{\mathrm{d}^2\sigma\left(\photonscatang, \eComptonenergy{}\right)}{\mathrm{d}\photonscatang \mathrm{d}\eComptonenergy{}} \ .\label{eq:twofoldsigma}
\end{align}

For valence-shell ionization, this corresponds to the standard approach of observing Compton scattering using a photon detector positioned at angle \photonscatang{}, measuring the photon energy \photonOutE{} $=$ \photonInE{} $-$ \Ebind{} $-$ \eComptonenergy. For inner-shell ionization followed by Auger decay, this connection between photon energy loss and the Compton electron energy is less straightforward, since Auger and Compton electrons can exchange energy via post-collision interaction, as we will discuss later.

Figures~\ref{Fig4} and \ref{Fig5} show the doubly differential cross section [Eq.~(\ref{eq:twofoldsigma})] for Compton scattering at the Ne L (2s and 2p) and K shells, respectively. Panels (a) of both figures display the relative \eCompton{} intensity as a function of the Compton electron energy \eComptonenergy{} and the scattering angle \photonscatang{}, revealing dramatic differences between the two cases. The only discernible differences between the data in Figs.~\ref{Fig4}(a) and \ref{Fig5}(a) are the orbital from which the Compton electron originates and the presence of an Auger electron. This provides answers to the question posted in the introduction: What effect does the binding of the scattering electron have on the kinematics? To answer this, we compare the shape of the energy distributions in Figs.~\ref{Fig4}(a) and \ref{Fig5}(a) to the black solid curves in the same panels, given by

\begin{align}
    \eComptonEbound(\photonscatang) = \eComptonEfree(\photonscatang) - \Ebind \ . \label{eq:Eebound}
\end{align}

That is, the Compton electron energy is given by the energy transfer to a free electron at rest minus the binding energy of the electron's orbital.

We first discuss Compton scattering at the Ne L shell (Fig.~\ref{Fig4}). The electron energy distribution clearly follows the black solid curve from Eq.~(\ref{eq:Eebound}), showing that for most scattering angles, the impulse approximation is valid. Within this approximation, the scattered photon energy distribution reflects the momentum distribution of the bound electron; other binding effects are neglected. This is emphasized by the dashed curves in Fig.~\ref{Fig4}. They
correspond to electrons that were initially not at rest. Instead, they had a momentum with a magnitude corresponding to the full width half maximum of the initial electron momentum distribution of the Ne L shell \cite{Daoud_1985} along or opposite to the direction of the momentum transfer
\MomTransVec{}. The width of the electron energy distribution in Fig.~\ref{Fig4}(a) clearly coincides with these dashed curves. Projections of the electron energy for different scattering angles are shown in Figs.~\ref{Fig4}(b-d), highlighting the systematic behavior of the distribution. For small scattering angles [Fig.~\ref{Fig4}(b)], the energy transfer in the Compton scattering process is so small that an electron initially at rest would not overcome the ionization threshold \Ebind{}. For electrons to overcome \Ebind{} upon a momentum boost Q,
they must have some initial momentum, resulting in a final non-zero kinetic energy. Accordingly, for small scattering angles, only the high-momentum tail of the initial-state wave function contributes to ionization, leading to low-energy continuum electrons. With increasing photon scattering angles, the full initial-state momentum distribution is lifted to the continuum, producing a broad energy distribution, as seen in Figs.~\ref{Fig4}(c,d). These results already point to the explanation of the dramatic differences that are seen in the
corresponding distributions for Ne K-shell Compton scattering (Fig.~\ref{Fig5}), which we discuss next.

A significant portion of scattering angles result in calculated negative energy as given by
$\eComptonEbound = \eComptonEfree - 0.87$~keV [Eq.~(\ref{eq:Eebound})]. Hence, for K-shell electrons, the observed energy distribution differs drastically from \eComptonEbound{}, and the na\"ive quasi-free electron prediction fails. Nonetheless, the maximum of the low-energy electron distribution occurs at the scattering angle where \eComptonEbound{} becomes positive [black solid curve in Fig.~\ref{Fig5}(a)]. A peak of low-energy electrons is always present, independent of scattering angle, confirming the predictions in Ref.~\cite{Bergstrom1993}, where a peak structure in the electron energy distribution with increasing scattering
angle was discussed. This low-energy peak persists even when $\eComptonEbound \gg 0$. One possible origin is post-collision interaction (PCI) \cite{Niehaus1977} between Compton and subsequently emitted Auger electrons [Eq.~(\ref{eqK2})], a well-known effect in photoionization processes. Low-energy Compton electrons experience a dramatic change in the Coulomb potential from $Z=1$ to $Z=2$ upon Auger-Meitner decay, as the (potentially significantly) faster Auger electron overtakes the Compton electron. The Compton electron loses energy while the Auger electron gains energy. The scattered photon remains unaffected by PCI, explaining why traditional photon-in-photon-out experiments did not observe this effect. As shown in Ref.~\cite{Melzer2024}, this is not the only source of low-energy electrons, since they also appear in L-shell Compton scattering. Unlike L-shell ionization, the K-shell electron energy distribution cannot be described well by \eComptonEfree, even if $\eComptonEfree \gg I_P$. To further explore this, we now discuss higher-order differential observables.

\section{Three-fold differential cross sections}
\label{secthreefold}

Comparing Figs.~\ref{Fig4}(a) and \ref{Fig5}(a) shows dramatic differences, which can be attributed to the breakdown of the impulse approximation for Ne K-shell ionization. However, for large scattering angles, the energy transfer in the Compton process is much larger than the binding energy even of the Ne K shell. As an example, consider a scattering angle that corresponds to a final electron energy given by \eComptonEbound{} [Eq.~(\ref{eq:Eebound})] of 0.5~keV (72.1$^\circ$ for Ne L and 151.4$^\circ$ for Ne K). For Compton scattering at Ne L, the measured electron energy distribution matches the prediction of \eComptonEbound{} quite well, while no such resemblance can be claimed for Ne K. To gain more insight into the underlying kinematics explaining this discrepancy, we will discuss triply differential cross sections in Cartesian coordinates [Eq.~(\ref{eqnfourcrossseckart}) but integrated over one electron momentum coordinate].

\begin{figure}
\includegraphics{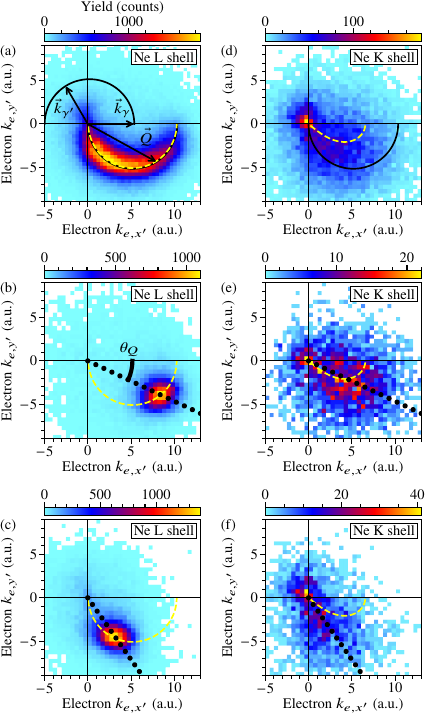}
\caption{Electron emission pattern for (a-c) Ne L-shell and (d-f) Ne K-shell ionization by Compton scattering of a 20~keV photon. Coordinate frame as shown in Fig.~\ref{Fig1}, i.e., Eqs.~(\ref{eq:coord}). The data are integrated over the electron momentum component out of the plane (coordinate $k_{e,z'}$). The black curve shows the momentum transfer for scattering at a free electron at rest. The yellow dashed curve shows the binding-energy-corrected momentum transfer at an electron at rest. (a,d) Data are integrated over all angles $\theta_{Q}$ of momentum transfer. (b,e) Subsets of the data from (a,d) selected for $20^\circ < \theta_Q < 30^\circ$, indicated by the black dotted line. (c,f) Subsets of data from (a,d) selected for $50^\circ < \theta_Q < 60^\circ$, indicated by the black dotted line.}
\label{Fig6}
\end{figure}

As coordinate frame for the Cartesian coordinates, we chose an ``intrinsic'' Compton-scattering coordinate frame, as shown in Fig.~\ref{Fig1}. This frame differs from that in Fig.~\ref{Fig3}, which is fixed in the laboratory frame, i.e., the $x$ direction is given by the incoming photon direction and the $y$ direction by the photon polarization. For the intrinsic frame, the incoming photon is kept along the $x$ direction, while the outgoing photon always scatters into the upper half-plane, that is,

\begin{align}
    \begin{split}
        \hat{x}' &= \photonInVec / \photonInMag \ , \\
        \hat{z}' &= (\MomTransVec \times \hat{x}') / |\MomTransVec \times \hat{x}'| \ , \\
        \hat{y}' &= \hat{z}' \times \hat{x}'.
    \end{split} \label{eq:coord}
\end{align}

In this coordinate frame, the momentum transfer \MomTransVec{} lies in the $x', y'$ plane and points into the lower-right quadrant. Before inspecting triply differential cross sections, we first consider the doubly differential ones in this intrinsic coordinate frame [Figs.~\ref{Fig6}(a,d)]. The two panels show the same data as Figs.~\ref{Fig3}(a,d). Figure~\ref{Fig6}(a) also shows an example configuration of the momentum vectors \photonInVec, \photonOutVec, and \MomTransVec{} for a scattering angle of $\photonscatang = 120^\circ$. Both panels (a) and (d) display black solid and yellow dashed curves in the lower half. The black solid curve represents the momentum transfer $\MomTransVec = \photonInVec - \photonOutVec$, i.e., the final momentum that a free electron at rest would have after Compton scattering (the so-called Bethe ridge). The yellow dashed curve shows an \Ebind-corrected momentum transfer, where the magnitude of \MomTransVec{} is reduced by $\sqrt{2\Ebind}$ (negative values are discarded). These curves are analogous to the black solid curves in Figs.~\ref{Fig4}(a) and \ref{Fig5}(a), but shown in momentum space. For Compton scattering at the Ne L shell, the electron momentum distribution clearly follows the \Ebind-corrected prediction given by the yellow curve; for Ne K-shell ionization, this prediction is poor, consistent with Figs.~\ref{Fig4}(a) and \ref{Fig5}(a). However, the intrinsic coordinate frame reveals additional features.

We first examine Compton scattering at the Ne L shell [Fig.~\ref{Fig6}(a)]. As expected, most electrons are emitted along the Bethe ridge, with a width given by the initial-state Compton profile. In addition to this main contribution, two further features are visible, both attributed to electrons interacting with the residual ion. First, a broad base distribution (light blue) appears; second, a backward-directed distribution close to the origin (darker blue). The former is attributed to electrons scattered at the nucleus without significant energy loss and is discussed in more detail in Ref.~\cite{Melzer2024}. The latter consists of low-energy electrons that are back-reflected -- opposite to the initial momentum transfer and toward the incoming photon -- by the nucleus. We refer to this feature as the \textit{recoil peak}. The terminology is adapted from ($e,2e$) studies, where a similar peak is observed for small-momentum-transfer electron-impact collisions \cite{Coplan1994}. For very small momentum transfer, the binary and recoil contributions evolve into the dipole emission pattern of the photoelectric effect.

The corresponding data for Ne K-shell ionization are shown in Fig.~\ref{Fig6}(d). At $\Ebind(\text{Ne~K})= 870.2$~eV, the energy loss due to binding modifies the Bethe ridge considerably (compare the black solid and yellow dashed curves). The forward-directed electron distribution loosely follows the Bethe ridge. As discussed above, the width of the distribution at the Bethe ridge is given by the initial-state Compton profile, which is much broader for the Ne K shell than for the Ne L shell, thereby washing out the structure. This width corresponds to the width of the scattered-photon energy distribution and forms the basis of Compton profile measurements. It was already observed in Compton's pioneering work \cite{AHCompton1922} and first explained by DuMond \cite{DuMond1929}. Compared to Ne L-shell ionization, the recoil peak is dramatically enhanced. The electrons forming this peak were already visible in Fig.~\ref{Fig5} as a strong low-energy peak. Two effects contribute: first, K-shell electrons are located much closer to the nucleus than L-shell electrons, enhancing backscattering; second, energy loss during escape from the Coulomb potential increases the number of low-energy electrons (see Fig.~3 in Ref.~\cite{Melzer2024}).

The underlying scattering mechanisms become even clearer when moving from doubly to triply differential cross sections [Figs.~\ref{Fig6}(b,c,e,f)]. Here, the direction of the momentum transfer $\theta_Q$ is selected (black dotted line). The binary collision of an electron at rest in the initial state that loses $I_p$ upon its escape would lead to electrons located in momentum space at the intersection of the yellow dashed curve and the black dotted line. For L-shell ionization, a pronounced maximum is found at this momentum-space position predicted by the binding-energy-corrected binary encounter. The width of the distribution is given by the L-shell initial state Compton profile. In addition, light blue oval regions elongated along the momentum transfer are visible in Figs.~\ref{Fig6}(b,c). These events, previously observed in Ref.~\cite{Melzer2024}, result from electron scattering at the nucleus. A slight enhancement near the origin is also visible in Fig.~\ref{Fig6}(c). For K-shell ionization, the binding-energy correction is much larger (yellow curve), and the Compton profile is much broader. This is evident in Fig.~\ref{Fig6}(e), which shows a very broad maximum centered at the intersection of the yellow dashed curve and the black dotted line. This intersection moves toward the origin in Fig.~\ref{Fig6}(f). For both momentum transfers, K-shell ionization leads to a pronounced peak near the origin, caused by Coulomb focusing \cite{Melzer2024}. Related effects have also been observed in strong-field ionization \cite{Blaga09} and ion-atom collisions \cite{Reinhold1998}.

\section{Fully differential cross sections}
\label{secfourfold}

The data shown in Figs.~\ref{Fig6}(b,c,e,f) are not yet fully differential, as they are integrated over the electron momentum component out of the plane of the figure. Fully differential cross sections -- as argued in the introduction being four-fold differential in the present case -- are the most sensitive quantity for comparison with theory. One set of coordinates that has proven particularly useful for this purpose in ($e,2e$) and ion-impact studies are electron angular distributions for fixed electron energy and fixed magnitude of the momentum transfer with respect to the angle $\chi$ between the electron momentum and the photon momentum transfer \cite{Duerr2006, Ehrhardt1986, Fischer2003, Schulz2003}:

\begin{align}
\cos\chi = \frac{\eComptonVec \cdot \MomTransVec}{|\eComptonVec||\MomTransVec|} \ .
\end{align}

\begin{figure}[t]
\includegraphics{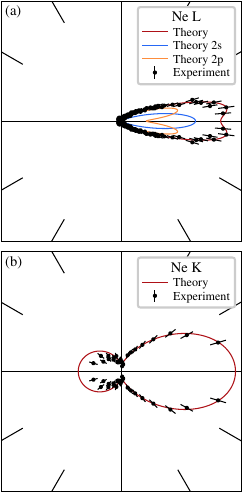}
\caption{Angular distributions of $\chi$ for Ne L- and Ne K-shell ionization in polar representation. (a) Angular distribution of Compton electrons for Ne L-shell ionization and, in addition, for 2s (blue curve) and 2p (orange curve), respectively. The momentum transfer \MomTransMag{} is restricted to $4.6 \pm 1.4$~a.u., the electron energy to $230 \pm 5$~eV. (b) Same as (a) but for Ne K-shell ionization. \MomTransMag{} is restricted to $4.6 \pm 1.4$~a.u., the electron energy to $230 \pm 87$~eV. Experimental data are shown as black dots and are normalized in (a,b) to the forward peak of the theory (red curve).
The error bars represent the standard statistical error.
}
\label{Fig7}
\end{figure}

\begin{figure}[t]
\includegraphics{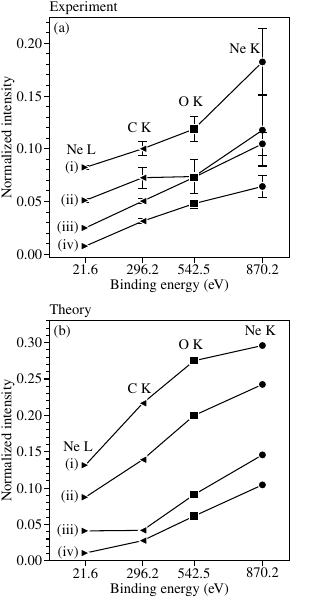}
\caption{Relative intensity of Compton electrons with $\chi\ge140^\circ$ for Ne L, C K, O K, and Ne K Compton ionization. (a) Experimental values are shown for different photon momentum transfers and electron energies, namely, from top to bottom: (i) $1.9 \pm 1.4$~a.u. and $39 \pm 35$~eV; (ii) $1.9 \pm 1.4$~a.u. and $124 \pm 50$~eV; (iii) $4.6 \pm 1.4$~a.u. and $230 \pm 87$~eV; (iv) $4.6 \pm 1.4$~a.u. and $403 \pm 87$~eV. (b) Theoretical backward emission calculated for the same main values and normalized to the integral.
The error bars represent the standard statistical error.
}
\label{Fig8}
\end{figure}

The electron angular distribution as a function of $\chi$ is shown in Figs.~\ref{Fig7}(a) and \ref{Fig7}(b) for Ne L- and Ne K-shell ionization, respectively. $\chi=0^\circ$ means the electron is emitted in direction of \MomTransVec{} and is pointing from left to right in Fig.~\ref{Fig7}. Here, the electron energy \eComptonenergy{} and the magnitude \MomTransMag{} are fixed (see figure caption). The angular distributions show two lobes, one in the direction of the momentum transfer and one opposite to it. In ($e,2e$) studies, the names binary and recoil peak have been coined for these features. Our fully differential experimental findings (black dots) show, again, a dramatic enhancement of the recoil peak for K-shell ionization, which is confirmed by our theoretical modeling in the $A^2$ approximation (red curves). These calculations were performed using the stationary single-center method \cite{SC1,SC2,SC3,SC4}, as described in detail in previous works on Compton ionization of atoms and molecules \cite{Melzer2024,Haubenreisser2025,Haubenreisser2025chiral}.
Our theory, based on the Hartree-Fock method, neglects electron-electron correlations. In addition, contributions of the $\vec p \cdot \vec A$ term of the light-matter interaction Hamiltonian was neglected. The $\vec p \cdot \vec A$ term contributes to features in the Compton spectra, such as infrared rise and resonant Raman scattering \cite{Bergstrom1993, Gavrila1972}, which are unobserved by our experimental setup.
Note that, in order to collect significant statistics, the electron energy range for K-shell ionization is significantly larger than for L-shell ionization.

A feature of note is observed in the distribution of $\chi$ for the Ne L-shell ionization. Figure~\ref{Fig7}(a) shows a dip at the very tip of the distribution (i.e., at $\chi=0^\circ$). Its origin becomes apparent when recalling the ramifications of the $A^2$ approximation. Within the approximation, the electron ground-state wave function in momentum space is shifted by \MomTransVec. The final-state wave function results from the projection of the momentum-boosted initial-state wave function onto the continuum scattering states. For large momentum transfers (compared to the extent of the initial-state momentum distribution), this is simply the momentum-boosted initial-state wave function, which forms the basis of the aforementioned impulse approximation. In momentum space, the tip of \MomTransVec{} therefore corresponds to momentum at $\chi=0^\circ$ in the initial-state momentum distribution. The momentum distribution of the Ne~2s orbital has a maximum at $\chi=0^\circ$. For the distribution of Ne~2p, there are three different contributions. The 2p$_{0}$ contribution has a maximum as well, while the 2p$_{-1}$ and 2p$_{+1}$ contributions both have a node. Therefore the Ne~2p momentum distribution in total shows a dip at $\chi=0^\circ$. This corresponds to the final angular distributions shown by the blue (2s) and orange (2p) curves in Fig.~\ref{Fig7}(a). Experimentally, 2s and 2p electrons cannot be distinguished; only their sum is measured. Nevertheless, traces of the dip in the 2p distribution are still visible in both the theoretical predictions (red curves) and the experimental observations (black dots).

To investigate the Compton recoil peak in more detail, we include, in addition to data for the Ne K and L shell, also data for C K- and O K-shell ionization. The intensity of back-scattered electrons relative to all electrons is shown in Fig.~\ref{Fig8}, both for the experimental results [panel (a)] and the theoretical modeling [panel (b)]. As noted above, the intensity of back-scattering for Ne K-shell ionization is larger than for Ne L-shell ionization. We therefore expect a dependence of the recoil peak on the binding energy and, furthermore, on the momentum transfer \MomTransMag{}. Theory and experiment show semi-quantitative agreement, and both confirm the expected dependence on the binding energy.
A possible source of the discrepancies is the lack of electron-electron correlations neglected in the theory.
When focusing on one of the datasets (i)-(iv) in Fig.~\ref{Fig8}, one can observe that, at fixed \MomTransMag{} and \eComptonenergy{}, increasing the binding energy leads to an increased intensity of back-scattered electrons. At the same time, fixing the binding energy while decreasing the momentum transfer also results in an increased backward intensity [cf. datasets (i, ii) and (iii, iv) for small and large momentum transfers, respectively].

This dependence on binding energy and momentum transfer arises from two effects. Classically, electron scattering increases with the charge of the scattering center and decreases with impact parameter and electron energy. The latter already explains the observed trend as a function of $Q$. For the K shell, the increase with target nuclear charge -- already present in the Rutherford cross section -- is clearly observed. The difference between the results for Ne L- and K-shell ionization mainly reflects the mean radius of these orbitals, which influences the most probable impact parameter.

\section{Conclusions}
\label{secconclusion}

We presented a comprehensive experimental and theoretical study of Compton scattering, providing insights on many levels of detail, from textbook-type overviews of the electron momentum emission pattern to the most detailed level of fully differential cross sections. Compton scattering is often described within the ``quasi-free'' electron approximation. In this scenario, the Compton-scattered electron is treated as unbound. Our study elucidates several breakdowns of this approximation. Two of them -- the role of the initial-state momentum distribution \cite{DuMond1929, Cooperbook} and a breakdown of the independent electron approximation \cite{Spielberger1999, Chuluunbaatar2021272, Chuluunbaatar2021278, Kaplan2003} -- have been extensively studied in the past. The present paper adds comprehensive data exploring two further consequences of electron binding, namely the influence of the binding energy of the Compton electron and its interaction with the ionic potential. Both result in dramatic changes in the electrons' kinematics. We measured Compton ionization of the Ne K, C K, O K, and Ne L shells and compared the experimental data with predictions of the $A^2$ approximation \cite{Pisk_2016,BERGSTROM19973}.

The energy distribution of Compton electrons from the Ne L shell demonstrates that, for low binding energies, the impulse approximation is still valid. In this case, the Compton electron's energy can be calculated by subtracting the electron's binding energy from the energy transfer to a free electron. For Ne K-shell ionization, additional strong effects are observed. Independent of the photon scattering angle, a low-energy peak appears in the electron energy distribution. This low-energy peak also persists under kinematical conditions where the energy transfer exceeds the binding energy. Low-energy electrons have also been detected for L-shell ionization.

Exploring the data in the photon-scattering coordinate frame, defined by the scattered photon and the momentum transfer, allowed us to understand this observation. Inspecting the electron momentum distribution in this intrinsic coordinate frame reveals scattering of the electron at its parent nucleus and a recoil peak of back-scattered electrons, as known from ($e,2e$) experiments. The intensity of this backscattering effect scales, as expected, with the size of the orbital, the momentum transfer, and the nuclear charge. It is this so far little-discussed interaction of the Compton electron with its parent ion, which makes Compton scattering a valuable tool for studying molecular potentials and for molecular imaging \cite{Haubenreisser2025,Haubenreisser2025chiral}.

\begin{acknowledgments}
This work was supported by the Deutsche Forschungsgemeinschaft (DFG, German Research Foundation).
We acknowledge the European Synchrotron Radiation Facility (ESRF) for the provision of synchrotron radiation facilities under proposal number CH-6396 (DOI: 10.15151/ESRF-ES-878668690), and we would like to thank V. Honkimäki, H. Isern, and F. Russello for assistance and support in using beamline ID31.
F.T. acknowledges funding by the DFG -- Project 509471550, Emmy Noether Programme.
K. L. thanks the Alexander von Humboldt Foundation.
N.M.N. and P.V.D. acknowledge funding by the DFG -- project no.~492619011 (DE 2366/6-2).
\end{acknowledgments}

\section*{Data Availability}
The data that support the findings of this article are not publicly available. The data are available from the authors upon reasonable request.

\end{document}